\documentclass[prd,showpacs,preprintnumbers,amsmath,amssymb,twocolumn]{revtex4-2}

\usepackage{graphicx} 
\usepackage{amsthm,amsmath,amssymb}
\usepackage{mathrsfs}
\usepackage{color}
\usepackage{appendix}
\usepackage{hyperref}
\usepackage{booktabs} 
\usepackage{multirow}
\usepackage{makecell}
\usepackage{tabularx}
\usepackage{array}
\usepackage[normalem]{ulem}
\definecolor{mydeepgreen}{RGB}{0,110,30} 

\date{\today}
\begin{document}
\title{Searching for Solar-Basin Axionlike-Particle Decay with XMM-Newton Blank-Sky Observations}

\author{Bo Zhang}
\affiliation{Key Laboratory of Dark Matter and Space Astronomy, Purple Mountain Observatory, Chinese Academy of Sciences, Nanjing 210033, China}
\affiliation{School of Astronomy and Space Science, University of Science and Technology of China, Hefei 230026, China}

\author{Chi Zhang}
\affiliation{Key Laboratory of Dark Matter and Space Astronomy, Purple Mountain Observatory, Chinese Academy of Sciences, Nanjing 210033, China}
\affiliation{School of Astronomy and Space Science, University of Science and Technology of China, Hefei 230026, China}

\author{Lei Lei}
\affiliation{Key Laboratory of Dark Matter and Space Astronomy, Purple Mountain Observatory, Chinese Academy of Sciences, Nanjing 210033, China}
\affiliation{School of Astronomy and Space Science, University of Science and Technology of China, Hefei 230026, China}

\author{Yang Yu}
\affiliation{Key Laboratory of Dark Matter and Space Astronomy, Purple Mountain Observatory, Chinese Academy of Sciences, Nanjing 210033, China}
\affiliation{School of Astronomy and Space Science, University of Science and Technology of China, Hefei 230026, China}

\author{Guan-Shen Wang}
\affiliation{Key Laboratory of Dark Matter and Space Astronomy, Purple Mountain Observatory, Chinese Academy of Sciences, Nanjing 210033, China}
\affiliation{School of Astronomy and Space Science, University of Science and Technology of China, Hefei 230026, China}

\author{Bing-Yu Su}
\affiliation{Key Laboratory of Dark Matter and Space Astronomy, Purple Mountain Observatory, Chinese Academy of Sciences, Nanjing 210033, China}
\affiliation{School of Astronomy and Space Science, University of Science and Technology of China, Hefei 230026, China}

\author{Lei Feng}
\email{fenglei@pmo.ac.cn}
\affiliation{Key Laboratory of Dark Matter and Space Astronomy, Purple Mountain Observatory, Chinese Academy of Sciences, Nanjing 210033, China}

\begin{abstract}
Axion-like particles (ALPs) bound in the solar gravitational field form the so-called ALP solar-basin. Since the two-photon decay of non-relativistic particles is approximately isotropic, this population can be searched for using observations in the anti-solar direction. In this work, we propose a search strategy for narrow decay-line signals from the ALP solar basin using \textit{XMM-Newton} blank-sky observations (XMM-BSOs) stacked spectra data taken in directions opposite to the Sun. By jointly fitting the signal and background model, we obtain limits on $g_{a\gamma\gamma}^{95}$ in the mass range $m_a=1.4\text{--}16~{\rm keV}$, with typical sensitivities of $g_{a\gamma\gamma}\sim10^{-10}\text{--}10^{-11}~{\rm GeV}^{-1}$. We have implemented the first anti-solar search for the solar basin, demonstrating that this strategy can exploit the stacked exposure of a large number of X-ray observations and provide a scalable analysis framework for future searches.

\end{abstract}
\maketitle

\section{Introduction}

Axion-like particles (ALPs) are among the most widely studied weakly coupled particle candidates in physics beyond the Standard Model. Generic ALPs can arise in a variety of high-energy theoretical frameworks, including string theory\cite{Svrcek:2006yi,Arvanitaki:2009fg,Benabou:2026jtv,Benabou:2025kgx}, extra dimensions\cite{Dienes:1999gw}, and dark sectors\cite{Jaeckel:2010ni}, and their masses and couplings can span a broad parameter space \cite{Jaeckel:2010ni,Irastorza:2018dyq}. Motivated by their possible roles in addressing dark matter\cite{Marsh:2015xka}, baryogenesis or baryon asymmetry generation\cite{Co:2019wyp}, and naturalness problems\cite{Peccei:1977hh,Weinberg:1977ma,Wilczek:1977pj}, ALPs have become an important target of both laboratory searches \cite{CAST:2024eil, Ortiz:2020tgs, IAXO:2019mpb, QUAX:2024fut, ADMX:2025vom, Salemi:2021gck, Pandey:2024dcd, Adair:2022rtw, Arza:2021ekq, Redondo:2010dp, Ehret:2010mh, Bahre:2013ywa, Friel:2024shg, Isleif:2022ytq,Pappas:2025zld} and astrophysical probes \cite{Sisk-Reynes:2022sqd, HESS:2013udx, Depta:2020wmr, Langhoff:2022bij, Dev:2023hax, Diamond:2023cto, Goldstein:2024mfp, Noordhuis:2022ljw, Dessert:2021bkv, Dessert:2020lil, Dolan:2021rya, Hooper:2007bq, Fermi-LAT:2016nkz, Xia:2018xbt, Liang:2018mqm, Xia:2019yud, Zhang:2018wpc, Majumdar:2018sbv, Chen:2019fsq, Meyer:2020vzy, Yuan:2020xui, Li:2020pcn, Xiao:2020pra,Ning:2024ozs,Ning:2026ebu,Ning:2025kyu} over the past decade.

In this work, we focus on the interaction between ALPs and photons. The corresponding effective Lagrangian can be written as\cite{Peccei:1977hh,Weinberg:1977ma,Wilczek:1977pj,Bastero-Gil:2021oky}
\begin{equation}
\mathcal{L}
=
\frac{1}{2}\partial^\mu a\,\partial_\mu a
-
\frac{1}{2}m_a^2 a^2
-
\frac{1}{4}g_{a\gamma\gamma}a F^{\mu\nu}\tilde{F}_{\mu\nu},
\end{equation}
where $a$ denotes the ALP field, $m_a$ is the ALP mass, $F^{\mu\nu}$ is the electromagnetic field-strength tensor, $\tilde{F}_{\mu\nu}$ is its dual tensor, and $g_{a\gamma\gamma}$ is the ALP--photon coupling. With this interaction, ALPs can be produced in plasmas through the Primakoff process \cite{Primakoff:1951pj,Irastorza:2018dyq,Wu:2024fsf} and photon coalescence \cite{Redondo:2013wwa,Beaufort:2023zuj}, and they can be indirectly searched for through ALP--photon conversion in magnetic fields \cite{Sikivie:1983ip,Irastorza:2018dyq} or through their two-photon decay \cite{Cadamuro:2011fd,Beaufort:2023zuj}.
Hot stellar plasmas provide one of the most important astrophysical environments for ALP production\cite{Nguyen:2023czp}. Existing studies have used stellar systems over a wide range of scales as ALP sources, including the Sun in helioscope searches such as CAST and IAXO \cite{CAST:2017uph,IAXO:2019mpb}, nearby stellar systems such as Alpha Centauri with X-ray observations by \textit{Chandra} and \textit{eROSITA} \cite{Chen:2024ekh}, globular clusters through stellar-evolution constraints \cite{Ayala:2014pea,Dolan:2022kul}, and external galaxies such as M82 and M87 with \textit{NuSTAR} observations \cite{Ning:2024eky,Candon:2024eah}. 

In the keV mass range, particularly, the gravitational potential of a star can trap a fraction of the ALP phase space, leading to the formation of a gravitationally bound axion population around the star. For the Sun, this population is commonly referred to as the solar basin \cite{VanTilburg:2020jvl,DeRocco:2022jyq}. This mechanism exploits the accumulation of ALPs over the long lifetime of the star and can therefore lead to very stringent constraints on ALP couplings \cite{DeRocco:2022jyq,Beaufort:2023zuj,Chen:2024ekh}.

\begin{figure}
    \centering
    \includegraphics[width=1.0\linewidth]{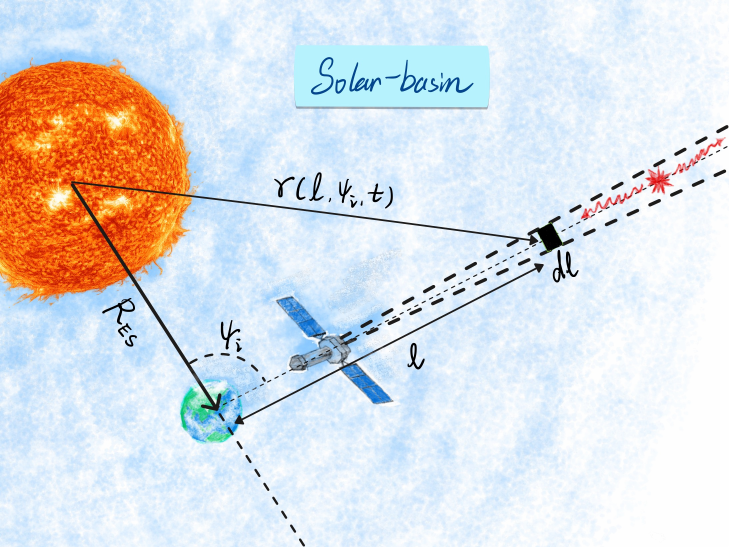}
    \caption{Schematic overview of the solar-basin ALP search strategy and its observational geometry. ALPs produced in the solar plasma can become gravitationally bound to the Sun and accumulate to form the solar basin. Non-relativistic ALPs decay approximately isotropically into two photons, allowing their decay signal to be searched for in the anti-solar direction. A large number of \textit{XMM-Newton} blank-sky observations are then stacked to search for the resulting narrow ALP decay-line signal.}
    \label{fig_schematic_diagram}
\end{figure}

The two-body decay of non-relativistic ALPs is approximately isotropic \cite{DeRocco:2022jyq,Qiu:2026irb}, hence the decay signal from the solar basin can be searched for using observations taken in directions opposite to the Sun as show in Fig.~\ref{fig_schematic_diagram}. In this work, we search for the two-photon decay signal from the ALP solar basin using anti-solar observations of \textit{XMM-Newton} . The public \textit{XMM-Newton} blank-sky observations (XMM-BSOs) data \cite{Foster:2021ngm,Dessert:2018qih} set stacks a large number of non-solar \textit{XMM-Newton} observations according to sky region and instrument/ring, providing a large total exposure and X-ray spectra well suited for narrow-line searches. We model the expected ALP decay signal from the solar basin and derive the corresponding count templates for the XMM-BSOs stacked spectra. We then search for the ALP decay signal in the XMM-BSO spectra by jointly fitting the signal templates and the constructed background model. Finally, we derive constraints on the ALP--photon coupling $g_{a\gamma\gamma}$.

This paper is organized as follows. Sec.~\ref{section_ALP_signal_model} describes the modeling of the anti-solar solar-basin ALP decay signal. Sec.~\ref{section_XMM_BSO} introduces the XMM-BSOs data products and derives the count templates applicable to the XMM-BSOs stacked spectra. Sec.~\ref{section_bacground_model} constructs the spectral background model used to fit the XMM-BSOs spectra. Sec.~\ref{section_result} introduces the profile-likelihood framework, presents the resulting constraints on $g_{a\gamma\gamma}$, and discusses their interpretation and robustness. Finally, we conclude in Sec.~\ref{conclusion}.

\section{Solar-Basin ALP Signal Modeling for XMM-BSO}
\label{section_ALP_signal_model}

\subsection{Density profile of solar-basin}
\label{subsection_solar_basin}

In this work, we consider ALPs coupled only to photons. In the plasma of the solar interior, ALPs can be produced through photon coalescence and the Primakoff process . The Primakoff process corresponds to the conversion of thermal photons into ALPs in the Coulomb fields of charged particles in the plasma. Since the velocities of gravitationally bound solar ALPs are much smaller than the speed of light, the Primakoff production rate $\Gamma_{\rm Prim}$ can be expanded in the small momentum. Keeping the leading non-vanishing term, the photon-to-ALP conversion rate is \cite{Raffelt:1985nk,DiLella:2000dn,Bastero-Gil:2021oky}
\begin{equation}
\Gamma_{\rm Prim}
\simeq
\frac{
g_{a\gamma\gamma}^2 T \kappa^2
}{
32\pi^2
}
\left[
\frac{
8p^2
}{
3\left(\kappa^2+m_a^2\right)
}
+
\mathcal{O}(p^4)
\right],
\end{equation}
where $p$ is the momentum of the produced ALP, $T$ is the local plasma temperature, and $\kappa$ is the Debye--Hückel screening scale. The screening scale is determined by the charged particles in the plasma,
\begin{equation}
\kappa^2
=
4\pi\alpha
\sum_{i=e,{\rm ions}}
\frac{Z_i^2 n_i}{T},
\end{equation}
where $\alpha$ is the fine-structure constant, and $Z_i$ and $n_i$ are the charge number and number density of the $i$-th charged species, respectively. The solar temperature, density, electron number density, and nuclear abundances used in our calculation are taken from the standard solar model \cite{Vinyoles:2016djt}.

Photon coalescence refers to the inverse decay process in which two thermal photons merge into one ALP, $\gamma+\gamma\to a$. Its production rate can be obtained from the ALP two-photon decay width through detailed balance \cite{Cadamuro:2010cz,Cadamuro:2011fd},
\begin{equation}
\scalebox{1.0}{$\Gamma_{\rm Coal}(T)
=
\Gamma_{a\gamma\gamma}
\frac{
m_a^2-4\omega_p^2
}{
m_a^2
}
\frac{m_a}{E_a}
\left[
1+
\frac{2T}{p}
\ln
\left(
\frac{
1-e^{-(E_a+p)/(2T)}
}{
1-e^{-(E_a-p)/(2T)}
}
\right)
\right]$}
,
\end{equation}
where $E_a$ is the ALP energy, $\omega_p=\sqrt{4\pi\alpha n_e/m_e}$ is the plasma frequency, and $\Gamma_{a\gamma\gamma}$ is the ALP vacuum two-photon decay width,
\begin{equation}
\Gamma_{a\gamma\gamma}
=
\frac{
g_{a\gamma\gamma}^2 m_a^3
}{
64\pi
}.
\end{equation}
The local ALP production rate in the solar interior is then given by
\begin{equation}
\Gamma_{\rm Prod} =
\Gamma_{\rm Prim}
+
\Gamma_{\rm Coal}.
\end{equation}
It is worth noting that the Primakoff process is suppressed by $p^2$ for non-relativistic bound ALPs\cite{DiLella:2000dn}. By contrast, photon coalescence is subject to the plasma threshold suppression when $m_a \leq 2\omega_p$. For heavier ALPs, photon coalescence can become more efficient than the Primakoff process\cite{Chen:2024ekh}.

If ALPs produced in the solar interior have velocities below the local escape velocity, they remain bound in the solar gravitational field. Over the lifetime of the Sun, these bound ALPs accumulate in the gravitational potential and form a gravitationally bound ALP population around the Sun, referred to as the solar basin. In a spherically symmetric solar model and in the approximation far from the solar surface, the number density of the gravitationally trapped ALP population can be written as \cite{DeRocco:2022jyq,Beaufort:2023zuj}

\begin{widetext}
\begin{equation}
n_a(t_\odot,\bar r)
=
\frac{
1-e^{-t_\odot\Gamma_{a\gamma\gamma}}
}{
\Gamma_{a\gamma\gamma}
}
\,
\frac{1}{4\pi \bar r^4}
\int_0^1 d\bar r_0\,
\bar r_0^2\,
\mathcal{S}(\bar r_0,m_a)
\sqrt{
2\left[
\bar\Phi(\bar r_0)
-
\frac{1}{\bar r}
\right]
},
\end{equation}
\end{widetext}

where $\bar r=r/R_\odot$ is the radial position normalized by the solar radius, $\bar r_0$ is the ALP production position inside the Sun, and $\bar\Phi(\bar r_0)$ is the dimensionless solar gravitational potential. We define the local ALP source term as
\begin{equation}
\mathcal{S}(\bar r_0,m_a)
\equiv
\int_{\rm bound}
\frac{d^3p}{(2\pi)^3}\,
\Gamma_{\rm Prod}(\bar r_0,p)\,
\frac{1}{e^{E_a/T(\bar r_0)}-1},
\end{equation}
where the integration domain is determined by the bound-orbit condition $v_a=p/E_a<v_{\rm esc}(\bar r_0)$. The factor $1-e^{-t_\odot\Gamma_{a\gamma\gamma}}$ describes the accumulation of bound ALPs with a finite lifetime: when $\Gamma_{a\gamma\gamma}t_\odot\ll1$, the number density grows approximately linearly with time, whereas when $\Gamma_{a\gamma\gamma}t_\odot\gtrsim1$, the bound-state density gradually approaches decay equilibrium. In this work, we take the solar evolution time to be $t_\odot = 5 {\text{ Gyr}}$.

\subsection{Decay Flux from Stacked Anti-Solar Observations}
\label{subsection_observation}

A stellar axion basin can be constrained through the two-photon decay of gravitationally bound ALPs. Since the solar gravitational potential is shallow, the trapped ALPs must originate from the low-energy tail of the production spectrum. As a result, the bound ALPs in the solar basin are non-relativistic.

In the satellite rest frame, the two-photon decay is approximately isotropic for non-relativistic ALPs. This implies that the decay signal from the solar basin is not confined to the direction toward the Sun. As long as the detector line of sight passes through the ALP cloud, decay photons can be received by the detector. Therefore, in principle, all non-solar deep-sky exposures of X-ray satellites can serve as valid observations for probing the decay of the solar-basin ALP population.

To fully exploit X-ray satellite observations in constraining the ALP decay signal, one needs to account for the fact that real X-ray data consist of many independent observations. Each observation has its own ALP signal template, determined mainly by the solar viewing angle, exposure time, field of view, and detector response. When interpreting the stacked data, the ALP decay signal should therefore be computed separately for each observation and then stacked in the same way as the observational data.

The geometry of the anti-solar solar-basin ALP decay search with \textit{XMM-Newton} blank-sky observations is illustrated schematically in Fig.~\ref{fig_schematic_diagram}. We now derive the time-integrated incident photon flux spectrum from ALP decays within the field of view of a single observation. The photon production rate per unit volume from ALP decay is
\begin{equation}
\frac{dR}{dV}
=
2\Gamma_{a\gamma\gamma}n_a(r),
\end{equation}
where the factor of two accounts for the two photons produced in each decay. Assuming that the decay is isotropic in the ALP rest frame, the emissivity per unit volume, per unit solid angle, and per unit energy is
\begin{equation}
j_{\gamma}(E,r)
=
\frac{1}{4\pi}\frac{dR}{dV}
\delta\left(E-\frac{m_a}{2}\right)
=
\frac{2\Gamma_{a\gamma\gamma} n_a(r)}{4\pi}
\delta\left(E-\frac{m_a}{2}\right).
\end{equation}
The photon intensity observed by the satellite along a direction $\hat n$ is then
\begin{align}
I_{\gamma}(E,\hat n,t)
&=
\int_0^\infty d\ell\,
j_{\gamma}\!\left(E,r(\ell,\psi_i,t)\right) \nonumber \\
&=
\frac{\Gamma_{a\gamma\gamma}}{2\pi}
\int_0^\infty d\ell\,
n_a\!\left[r(\ell,\psi_i,t)\right]
\delta\left(E-\frac{m_a}{2}\right),
\end{align}
where $\ell$ is the line-of-sight integration variable. The distance between the line-of-sight point and the solar center is
\begin{equation}
r(\ell,\psi_i,t)
=
\sqrt{
R_{\rm sat}^2(t)
+\ell^2
-2R_{\rm sat}(t)\ell\cos\psi_i(t)
},
\end{equation}
where $R_{\rm sat}(t)$ is the distance between the satellite and the Sun, and $\psi_i(t)$ is the solar viewing angle of the observation. For the $i$-th observation, with start time $t_i$, duration $\Delta t_i$, and field of view $\Delta\Omega_i$, the time-integrated incident photon flux spectrum is

\begin{widetext}
\begin{equation}
\mathcal{F}_{\gamma,i}(E)
=
\frac{\Gamma_{a\gamma\gamma}}{2\pi}
\int_{t_i}^{t_i+\Delta t_i} dt
\int_{\Delta\Omega_i} d\Omega
\int_0^\infty d\ell\,
n_a\!\left[r(\ell,\psi_i,t)\right]
\delta\left(E-\frac{m_a}{2}\right).
\end{equation}
\end{widetext}
After the photons arrive at the detector, the observed counts are obtained by convolving this incident flux with the instrumental response, including the effective area, field-of-view response, and energy redistribution.

For the specific case considered in this work, we adopt the following three approximations to simplify the calculation:
\begin{enumerate}
    \item \textit{Satellite-position approximation.}
    Since the satellite orbit is much smaller than the Earth--Sun distance, we set
    \begin{equation}
    R_{\rm sat}(t)=R_{\rm ES}=1~{\rm AU}.
    \end{equation}

    \item \textit{Small-field-of-view approximation.}
    The field of view of the satellite is small. We therefore approximate the angular integral over the field of view by the product of the field-of-view solid angle and the line-of-sight integral along the central pointing direction,
    \begin{equation}
    \int_{\Delta\Omega_i} d\Omega\, J(\hat n,t)
    \simeq
    \Delta\Omega_i\,J(\hat n_i,t).
    \end{equation}

    \item \textit{Fixed-geometry approximation during one observation.}
    The duration of a single X-ray observation is typically shorter than the time scale over which the solar direction changes significantly on the sky. We therefore treat the solar viewing angle and the line-of-sight geometry as constant during one observation, taking their values at the midpoint of the exposure,
    \begin{equation}
    \int_{t_i}^{t_i+\Delta t_i} dt\, J_i(t)
    \simeq
    \Delta t_i\,J_i(t_i+\Delta t_i/2),
    \end{equation}
    where $J_i(t)$ denotes the relevant time-dependent geometric factor schematically.
\end{enumerate}
Combining these approximations, the time-integrated incident photon flux spectrum from ALP decay during the $i$-th observation can be written as
\begin{equation}
\mathcal{F}_{\gamma,i}(E)
\simeq
\frac{\Gamma_{a\gamma\gamma}}{2\pi}
\Delta t_i
\Delta\Omega_i
J(\psi_i)
\delta\!\left(E-\frac{m_a}{2}\right),
\end{equation}
where
\begin{equation}
J(\psi_i)
=
\int_0^\infty d\ell\,
n_a\!\left[
\sqrt{
R_{ES}^2
+\ell^2
-2R_{ES}\,\ell\cos\psi_i
}
\right].
\end{equation}
If the final X-ray data set is obtained by stacking multiple independent observations, the corresponding ALP decay signal template should be constructed as
\begin{equation}
\mathcal{F}_{\gamma}^{\rm stack}(E)
=
\sum_i
\mathcal{F}_{\gamma,i}(E).
\end{equation}

For comparison with the actual X-ray data, this incident photon signal must further be matched to the detector effective area, the energy response function, and the stacking procedure used to construct the data set.
\newline
\subsection{Detector Response and Count-Template Construction}
\label{subsection_decay_line}

The previous subsection derived the time-integrated incident photon flux spectrum from ALP decays for stacked observations. To compare this signal with the actual X-ray data, the incident photons must be convolved with the detector response to construct the count template in the observed energy bins.

For the $i$-th observation, we denote by $A_i(E,\Omega,t)$ the detector effective area for a photon with true energy $E$ and incident direction $\Omega$ at time $t$, and by $R_i(E_{\rm obs}\mid E,\Omega,t)$ the response function for reconstructing this photon with observed energy $E_{\rm obs}$. For example, the probability that a photon with true energy $E$ is reconstructed in the observed energy bin $E_k\in [E_k^-,E_k^+]$ is
\begin{equation}
P_{i,k}(E,\Omega,t)
=
\int_{E_k^-}^{E_k^+}
dE_{\rm obs}\,
R_i(E_{\rm obs}\mid E,\Omega,t).
\end{equation}
The signal counts in the $E_k$ bin for a single observation are therefore
\begin{widetext}
\begin{equation}
S_{i,k}
=
\frac{\Gamma_{a\gamma\gamma}}{2\pi}
\int_{t_i}^{t_i+\Delta t_i}dt
\int_{\Delta\Omega_i}d\Omega
\int_0^\infty d\ell\,
n_a\!\left[r(\ell,\psi_i,t)\right]
A_i\!\left(\frac{m_a}{2},\Omega,t\right)
P_{i,k}\!\left(\frac{m_a}{2},\Omega,t\right).
\end{equation}
\end{widetext}
Using the same approximations as in the previous subsection, the count template can be simplified to
\begin{equation}
S_{i,k}
\simeq
\frac{\Gamma_{a\gamma\gamma}}{2\pi}
\Delta t_i\,
J(\psi_i)\,
\mathcal{T}_{i,k}\!\left(\frac{m_a}{2}\right),
\end{equation}
where we have defined the instrumental transfer matrix for this observation as
\begin{equation}
\mathcal{T}_{i,k}(E)
\equiv
\frac{1}{\Delta t_i}
\int_{t_i}^{t_i+\Delta t_i}dt
\int_{\Delta\Omega_i}d\Omega\,
A_i(E,\Omega,t)\,
P_{i,k}(E,\Omega,t).
\end{equation}
If the variations of the effective area and energy response across the field of view and during a single observation are further neglected, the instrumental transfer matrix reduces to
\begin{equation}
\mathcal{T}_{i,k}(E)
\simeq
\Delta\Omega_i\,
A_i(E)\,
P_{i,k}(E).
\end{equation}

To construct the signal model for a stacked X-ray data set, we compute the single-observation signal counts $S_{i,k}$ for each observation and then stack them following the same procedure used in the data processing. The total stacked signal template is therefore
\begin{equation}
S_k^{\rm stack}
=
\sum_i S_{i,k}.
\label{eq:xmmbso_signal_template}
\end{equation}

\section{XMM-BSO ring-stacked data products}
\label{section_XMM_BSO}

Foster et al.~\cite{Foster:2021ngm} constructed stacked all-sky spectra from \textit{XMM-Newton} observations to search for decaying dark matter signals in the Milky Way. They publicly released the processed XMM-BSO data set\footnote{\url{https://github.com/bsafdi/XMM_BSO_DATA}}. The XMM-BSO sample contains a full-sky blank-sky exposure of \(215~{\rm Ms}\) for MOS and \(57~{\rm Ms}\) for PN after the \(|b|\geq2^\circ\) sky selection and data-quality cuts. The data set stacks the MOS and PN observations separately into 30 rings according to the angular separation between the line of sight and the Galactic Center. It also provides, for each ring, the corresponding instrumental response and the observation IDs included in the stack.

For the solar-basin signal considered in this work, the ideal procedure would be to stack the observations with weights determined by the solar viewing angle of each exposure, thereby fully exploiting the angular distribution of the basin signal. However, the XMM-BSO data products have already been stacked according to the Galactic-Center angle, and the original event-level data cannot be reorganized within the public ring-stacked spectra. Nevertheless, because the ALP decay signal is a narrow spectral line, most of the discriminating power is retained in the energy dimension. The XMM-BSO data therefore allow us to obtain competitive constraints with a relatively low data-processing cost.

We need to construct ALP signal count templates that are matched to the observations entering each XMM-BSO stack. For a given instrument and ring, denoted by $(X,r)$, the ALP signal counts are
\begin{equation}
S_{X,r,k}
=
\sum_{i\in\mathcal{S}_{X,r}}
S_{i,k},
\end{equation}
where $\mathcal{S}_{X,r}$ denotes the set of observations included in this stack. The public XMM-BSO data set does not provide the detailed single-observation transfer matrices $\mathcal{T}_{i,k}$. Instead, it provides the total exposure $T_{X,r}$, the effective field-of-view weight $W_{X,r}$, and the stacked instrumental response $\mathcal{D}_{X,r,k}(E)$ for each ring, satisfying
\begin{equation}
T_{X,r}
\equiv
\sum_{i\in\mathcal{S}_{X,r}}
\Delta t_i,
\end{equation}
\begin{equation}
W_{X,r}
=
T_{X,r}\,
\Omega^{\rm eff}_{X,r},
\end{equation}
and
\begin{equation}
\mathcal D_{X,r,k}(E)
\equiv
\frac{
\displaystyle
\sum_{i\in\mathcal{S}_{X,r}}
\Delta t_i\,
\mathcal T_{i,k}(E)
}{
W_{X,r}
}.
\end{equation}
Using the definition of $\mathcal T_{i,k}(E)$ from the previous section, the instrumental factor $W_{X,r}\mathcal D_{X,r,k}(E)$ can be written as
\begin{widetext}
\begin{equation}
W_{X,r}\,
\mathcal D_{X,r,k}(E)
=
\sum_{i\in\mathcal{S}_{X,r}}
\int_{t_i}^{t_i+\Delta t_i}dt
\int_{\Delta\Omega_i}d\Omega\,
A_i(E,\Omega,t)\,
P_{i,k}(E,\Omega,t).
\end{equation}
\end{widetext}

For the ALP signal, the geometric factor $J(\psi_i)$ depends on the solar viewing angle of each exposure. Since the public XMM-BSO data provide only ring-level responses, we cannot reconstruct the correlation between $J(\psi_i)$ and the instrumental response of each individual exposure. We therefore approximate the exposure-by-exposure stacking by using an exposure-time-weighted average geometric factor within each ring,
\begin{equation}
\sum_{i\in\mathcal{S}_{X,r}}
\Delta t_i\,
J(\psi_i)\,
\mathcal{T}_{i,k}(E)
\;\simeq\;
\overline{J}_{X,r}
\sum_{i\in\mathcal{S}_{X,r}}
\Delta t_i\,
\mathcal{T}_{i,k}(E),
\end{equation}
with
\begin{equation}
\overline{J}_{X,r}
\equiv
\frac{
\displaystyle
\sum_{i\in\mathcal{S}_{X,r}}
\Delta t_i\,J(\psi_i)
}{
\displaystyle
\sum_{i\in\mathcal{S}_{X,r}}
\Delta t_i
}.
\end{equation}
The resulting XMM-BSO count template for the ALP signal is then
\begin{equation}
S_{X,r,k}(m_a,g_{a\gamma\gamma})
\simeq
\frac{\Gamma_{a\gamma\gamma}}{2\pi}
\,
\overline{J}_{X,r}
\,
W_{X,r}
\,
\mathcal{D}_{X,r,k}
\!\left(
\frac{m_a}{2}
\right).
\end{equation}

We note that the exposure lists released with the public XMM-BSO data set do not directly provide the observation times required to compute the solar viewing angles. In this work, we use the Observation IDs to retrieve the corresponding \textit{XMM-Newton} observation metadata, obtain the start and end times of each exposure, and compute the solar viewing angle $\psi_i$.

\section{Spectral modeling strategy}
\label{section_bacground_model}

The XMM-BSO data consist of ring-stacked spectra constructed from a large number of \textit{XMM-Newton} observations grouped by Galactic-Center angle. Their spectral backgrounds therefore span different observation epochs, sky regions, and instrumental conditions. As a result, the background is not associated with a single physical component, but includes contributions from the astrophysical X-ray background, unresolved point sources, particle backgrounds, instrumental continuum emission, and instrumental fluorescence lines. Such stacked spectra are difficult to model from first principles in a fully predictive way. Since the ALP decay signal considered in this work is a narrow spectral line, the background model must satisfy two requirements: it should be flexible enough to describe the broad-band continuum shape, while avoiding excessive freedom that could absorb a potential narrow-line signal.

Our background modeling follows the analysis strategy developed by Foster et al.~\cite{Foster:2021ngm} for the XMM-BSO data. In their work, Gaussian processes were used to describe broad-scale continuum mismodeling in the XMM-BSO spectra. We adopt the same idea of a Gaussian-process continuum model, and construct a minimal set of narrow-line background templates based on the residual structures identified in pre-fits for each energy band. This strategy is designed to describe the known and stable background features with a low-dimensional nuisance model, while preserving the stability of the narrow-line search.

\subsection{Background model}

Because a single background model would become inadequate over an excessively broad energy range, we divide the spectra into three energy bands: $0.7\text{--}1.3~\mathrm{keV}$, $1.3\text{--}2.5~\mathrm{keV}$, and $2.5\text{--}8.0~\mathrm{keV}$. In each band, we adopt a unified background model of the form ``smooth continuum + GP residual + line-library templates''. For instrument $X=\mathrm{MOS},\mathrm{PN}$, ring $r$, and energy bin $k$, the background model is written as
\begin{equation}
B_{X,r,k}
=
B^{\rm cont}_{X,r,k}
+
B^{\rm line}_{X,r,k},
\end{equation}
where $B^{\rm cont}_{X,r,k}$ describes the continuum background, while $B^{\rm line}_{X,r,k}$ represents narrow-line background components, mainly including atomic lines from the astrophysical X-ray background and fluorescence lines produced by instrumental materials. The baseline background models adopted in the three energy bands are summarized in Table~\ref{tab:bg_models}.

\begin{table*}[t]
\centering
\caption{Fiducial spectral background models used in the three photon-energy bands.}
\label{tab:bg_models}
\renewcommand{\arraystretch}{1.28}
\setlength{\tabcolsep}{5.0pt}
\begin{tabular}{c l c c c}
\toprule
Band & Component & Model & Template form & Nuisance parameters \\
\midrule

\multirow{3}{*}{\makecell{Low band\\ $0.7\text{--}1.3~{\rm keV}$}}
&
Continuum
&
Polynomial $+$ GP
&
$p_{X,r,k}+g_{X,r,k}$
&
\makecell{$c_{X,r,n}$ local;\\ $\sigma_E=0.3$ fixed}
\\
\cmidrule(lr){2-5}
&
Empirical lines
&
\makecell{Fixed empirical\\ line library}
&
$\displaystyle
\sum_{\ell\in \mathcal{L}_{\rm low}}
a_{X,r,\ell}L_{X,r,\ell,k}
$
&
$a_{X,r,\ell}$ local
\\
\cmidrule(lr){2-5}
&
Shift templates
&
\makecell{Local line-shape\\ correction}
&
$\displaystyle
\sum_{\ell\in \mathcal{L}_{\rm shift}}
b_{X,r,\ell}S_{X,r,\ell,k}
$
&
$b_{X,r,\ell}$ local
\\

\midrule

\multirow{3}{*}{\makecell{Mid band\\ $1.3\text{--}2.5~{\rm keV}$}}
&
Continuum
&
Polynomial $+$ GP
&
$p_{X,r,k}+g_{X,r,k}$
&
\makecell{$c_{X,r,n}$ local;\\ $\sigma_E=0.3$ fixed}
\\
\cmidrule(lr){2-5}
&
Sky lines
&
\makecell{Profile-linked\\ physical lines}
&
$\displaystyle
\sum_{\ell\in \mathcal{L}_{\rm phys}}
A_{X,\ell}P_{X,r,\ell}L_{X,r,\ell,k}
$
&
\makecell{$A_{X,\ell}$ linked;\\ $P_{X,r,\ell}$ fixed}
\\
\cmidrule(lr){2-5}
&
Al-K complex
&
\makecell{Main line $+$\\ curvature correction}
&
$\displaystyle
a^{\rm AlK}_{X,r}L^{\rm AlK}_{X,r,k}
+
b^{\rm AlK}_{X,r}C^{\rm AlK}_{X,r,k}
$
&
\makecell{$a^{\rm AlK}_{X,r}, b^{\rm AlK}_{X,r}$\\ local}
\\

\midrule

\multirow{2}{*}{\makecell{High band\\ $2.5\text{--}8.0~{\rm keV}$}}
&
Continuum
&
Polynomial $+$ GP
&
$p_{X,r,k}+g_{X,r,k}$
&
\makecell{$c_{X,r,n}$ local;\\ $\sigma_E=0.3$ fixed}
\\
\cmidrule(lr){2-5}
&
Fixed narrow lines
&
\makecell{Profile-linked\\ line atlas}
&
$\displaystyle
\sum_{\ell\in \mathcal{L}_{\rm high}}
A^{\rm high}_{X,\ell}
P^{\rm high}_{X,r,\ell}
L^{\rm high}_{X,r,\ell,k}
$
&
\makecell{$A^{\rm high}_{X,\ell}$ linked;\\ $P^{\rm high}_{X,r,\ell}$ fixed}
\\

\bottomrule
\end{tabular}
\end{table*}

Since we directly model the stacked count-rate spectra, we introduce a low-order energy-dependent mean function for each $(X,r)$ data set and use a Gaussian process to describe broad-scale residuals around this mean model:
\begin{equation}
B^{\rm cont}_{X,r}(E)
=
p_{X,r}(E)
+
g_{X,r}(E).
\end{equation}
The low-order mean function is taken to be
\begin{equation}
p_{X,r}(E)
=
\sum_{n=0}^{N_b}
c_{X,r,n}\,
\phi_n(E),
\end{equation}
where $c_{X,r,n}$ are nuisance coefficients. The basis functions $\phi_n(E)$ are chosen as polynomial functions of the rescaled energy variable,
\begin{equation}
\phi_n(E)
=
\left[
\frac{
2E-(E_{b,\min}+E_{b,\max})
}{
E_{b,\max}-E_{b,\min}
}
\right]^n.
\end{equation}
The Gaussian-process residual $g_{X,r}(E)$ is modeled as
\begin{equation}
g_{X,r}(E)
\sim
\mathcal{GP}
\left(
0,
K(E,E')
\right),
\end{equation}
with the non-stationary covariance kernel adopted by Foster et al.~\cite{Foster:2021ngm,Frate:2017mai},
\begin{equation}
K(E,E')
=
A_{\rm GP}
\exp\left[
-\frac{(E-E')^2}{2EE'\sigma_E^2}
\right].
\end{equation}
In our fiducial analysis, we fix $\sigma_E=0.3$. This kernel has a correlation length that scales with energy, making it suitable for describing continuum mismatch that varies slowly across the X-ray spectrum. In the fit, the continuum parameters are treated independently for each $(X,r)$ data set; that is, the continuum normalization and shape parameters are not shared among different rings or instruments. This avoids over-constraining the stacked backgrounds from different sky regions, observation epochs, and instrumental conditions to a single physical model.

The narrow-line background component is constructed from fixed line templates convolved with the instrumental response. We write a generic narrow-line component as
\begin{equation}
B^{\rm line}_{X,r,k}
=
\sum_{\ell\in\mathcal L_X}
a_{X,r,\ell}\,
L_{X,r,\ell,k},
\end{equation}
where $\mathcal L_X$ is the instrument- and band-dependent line library, $a_{X,r,\ell}$ is the nuisance parameter describing the intensity of the $\ell$-th line in the $(X,r)$ data set, and $L_{X,r,\ell,k}$ is the line template after energy binning and instrumental broadening. For a narrow line with central energy $E_\ell$, the template is defined as
\begin{equation}
L_{X,r,\ell,k}
=
\int_{\Delta E_k} dE_{\rm obs}\,
R_{X,r}(E_{\rm obs}\mid E_\ell),
\end{equation}
where $R_{X,r}(E_{\rm obs}\mid E)$ is the energy response for the corresponding instrument and ring. This expression represents the count-rate template produced in the $k$-th observed energy bin when an ideal monochromatic line with true energy $E_\ell$, i.e. $\delta(E-E_\ell)$, is propagated through the XMM-BSO ring-level response. In practice, this template is obtained by interpolating the instrumental response provided by XMM-BSO at $E_\ell$.

In addition to these generic narrow-line templates, some energy ranges require local shape-correction templates. The residual line-like structures in different energy bands have different origins, and therefore the construction of the corresponding local nuisance modes also differs from band to band. In Table~\ref{tab:bg_models}, \(S_{X,r,\ell,k}\) denotes the low-band shift templates for line-centroid shifts or asymmetric composite-line residuals, while \(C^{\rm AlK}_{X,r,k}\) denotes the local curvature correction for the mid-band Al-K complex. We describe these details separately for each energy band below in Appendix~\ref{sec:appendix_background_details}.

If an independent intensity nuisance parameter $a_{X,r,\ell}$ were introduced for every narrow line in every $(X,r)$ data set, the model would acquire substantial local freedom around the corresponding line energies. When the ALP decay line is close to a background line, the independently floating line intensities in different rings could adjust separately and project out a significant fraction of the ALP signal template. In this case, the sensitivity in the affected energy range would be driven mainly by the nuisance freedom of the model rather than by the statistical power of the data.

To reduce the freedom of the background model, we adopt a profile-linked parameterization. The main idea is to first determine the relative strengths of stable background lines across different rings from background-only fits, and then fix this ring profile during the ALP search while retaining only one overall amplitude parameter for each line. Specifically, in the background-only fits, we first allow the $\ell$-th physical line to have an independent local amplitude $a^{\rm pre}_{X,r,\ell}$ in each $(X,r)$ data set. These fitted amplitudes are then used to construct a fixed ring profile,
\begin{equation}
P_{X,r,\ell}
=
\frac{
a^{\rm pre}_{X,r,\ell}
}{
\left[
\sum_{r'}
\left(a^{\rm pre}_{X,r',\ell}\right)^2
\right]^{1/2}
}.
\end{equation}

The profile $P_{X,r,\ell}$ preserves the relative distribution of the $\ell$-th stable background line across different rings. In the subsequent ALP search, both $P_{X,r,\ell}$ and $L_{X,r,\ell,k}$ are kept fixed, and only the overall amplitude $A_{X,\ell}$ of each physical line is profiled:
\begin{equation}
B^{\rm link}_{X,r,k}
=
\sum_{\ell\in\mathcal L_{\rm link}}
A_{X,\ell}
P_{X,r,\ell}
L_{X,r,\ell,k}.
\end{equation}
Here $\mathcal L_{\rm link}$ denotes the set of physical lines in the corresponding energy band that are treated with the profile-linked prescription. This parameterization allows the astrophysical lines to have non-trivial distributions across rings, while avoiding fully independent line amplitudes for the same narrow line in different rings. It therefore reduces the excessive absorption of the ALP decay-line signal by background-line nuisance parameters.

\subsection{Validation of the combined background model}
\label{sec:bg_validation}

\begin{figure}
    \centering
    \includegraphics[width=1.0\linewidth]{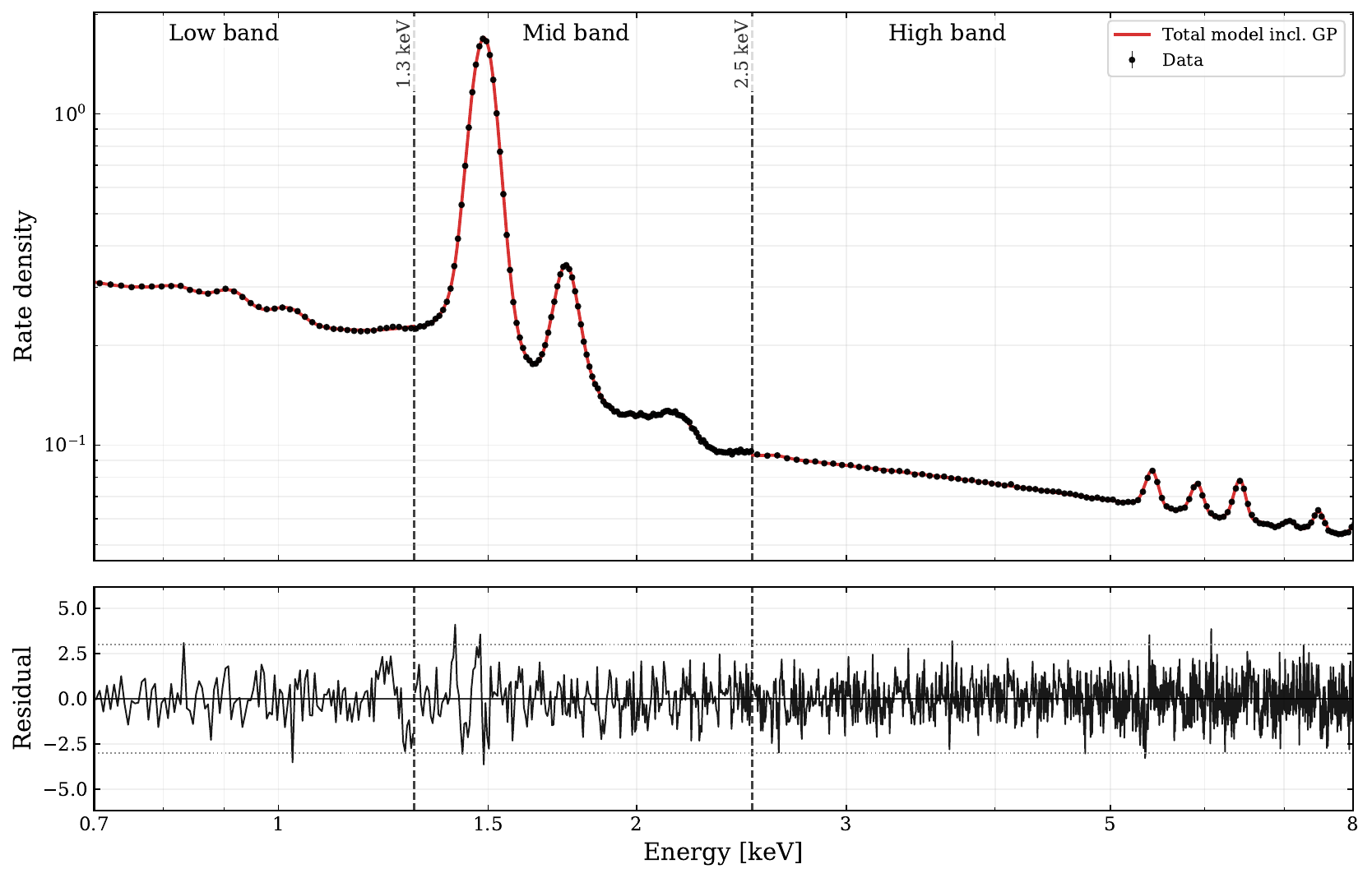}
    \caption{ Representative background fit to the XMM-BSO MOS ring-20 spectrum. The upper panel shows the rate-density spectrum over $0.7\text{--}8.0~{\rm keV}$, with the black vertical dashed lines marking the two band boundaries at $1.3~{\rm keV}$ and $2.5~{\rm keV}$. Black data points denote the observed spectrum and are mildly rebinned for visual clarity only; this rebinning is not used in the fit or in the statistical inference. The red solid curve shows the fiducial background model. The lower panel shows the whitened residuals computed using the original energy bins.}
    \label{fig:bg_fit_ring20}
\end{figure}

We fit the XMM-BSO spectra with the fiducial background models introduced above for the three energy bands. Fig.~\ref{fig:bg_fit_ring20} shows an example of the spectral fit for ring~20, with the black dashed vertical lines separating the energy ranges described by different background models. In the upper panels, the black points denote the observed data and the red solid curve gives the best-fit fiducial background model. The continuum component captures the broad spectral shape, while the inclusion of narrow-line templates allows the full fiducial model to reproduce the line-like structures in the data. The lower panels show the corresponding whitened residuals, which fluctuate around zero without a visually apparent systematic trend. We further quantify the stability of the background modeling below.

For each $(X,r)$ data set, we compute the whitened residuals of the best-fit background model. Denoting the observed spectrum by $d_{X,r,k}$, the best-fit background model by $\widehat B_{X,r,k}$, and the covariance matrix in energy band $b$ by $\mathbf C^{(b)}_{X,r}$, we define the residual vector as
\begin{equation}
\Delta_{X,r,k}
=
d_{X,r,k}
-
\widehat B_{X,r,k}.
\end{equation}
The corresponding whitened residual vector is
\begin{equation}
z^{(b)}_{X,r}
=
\left(\mathbf C^{(b)}_{X,r}\right)^{-1/2}
\Delta^{(b)}_{X,r}.
\end{equation}
After marginalizing over the Gaussian-process residual, the effect of the GP is equivalent to adding a GP covariance matrix to the statistical covariance of the data. Therefore, for the $(X,r)$ data set in energy band $b$, the total covariance is written as \cite{2015ITPAM..38..252A}
\begin{equation}
\mathbf C^{(b)}_{X,r;ij}
=
\sigma^2_{X,r,i}\delta_{ij}
+
\mathbf K^{(b)}_{X,r;ij},
\label{eq:cov_total}
\end{equation}
where $\sigma^2_{X,r,i}$ is the statistical variance of the $i$-th energy bin, and $\mathbf K^{(b)}_{X,r;ij}=K(E_i,E_j)$ is determined by the non-stationary GP kernel described above. If the background model provides an adequate description of the data and the covariance estimate is not strongly mismatched, the components of $z^{(b)}_{X,r}$ should be approximately distributed as random variables with zero mean and unit variance. We therefore use
\begin{equation}
\sigma_{\rm white}(X,r,b)
=
{\rm std}\!\left(z^{(b)}_{X,r}\right)
\end{equation}
as a diagnostic of the fit quality. We then perform a pooled whitened-residual test by combining the whitened residuals from all data sets.

\begin{figure*}[t]
\centering

\begin{minipage}[t]{0.49\textwidth}
\centering
\textbf{(a)}\\[-0.2em]
\includegraphics[width=\linewidth]{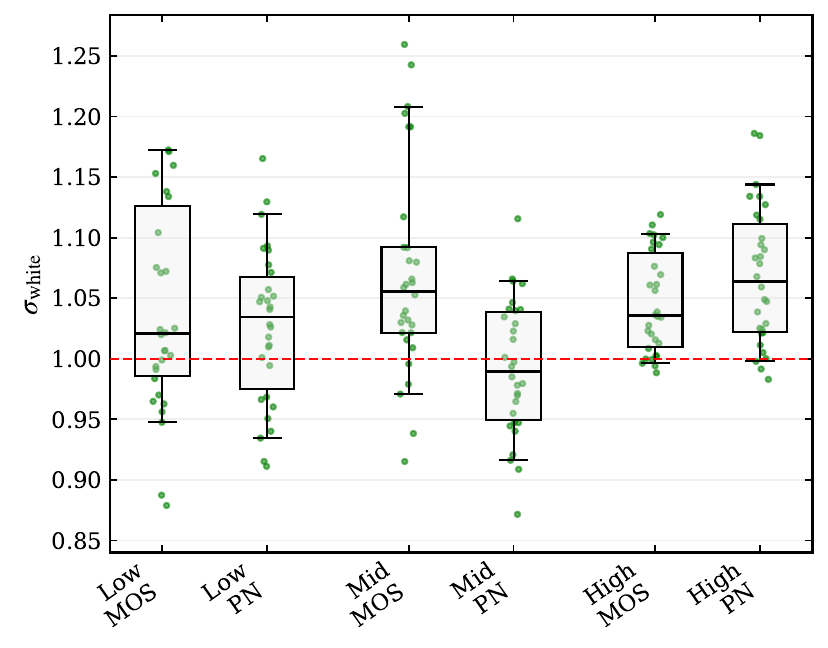}
\end{minipage}
\hfill
\begin{minipage}[t]{0.49\textwidth}
\centering
\textbf{(b)}\\[-0.2em]
\includegraphics[width=\linewidth]{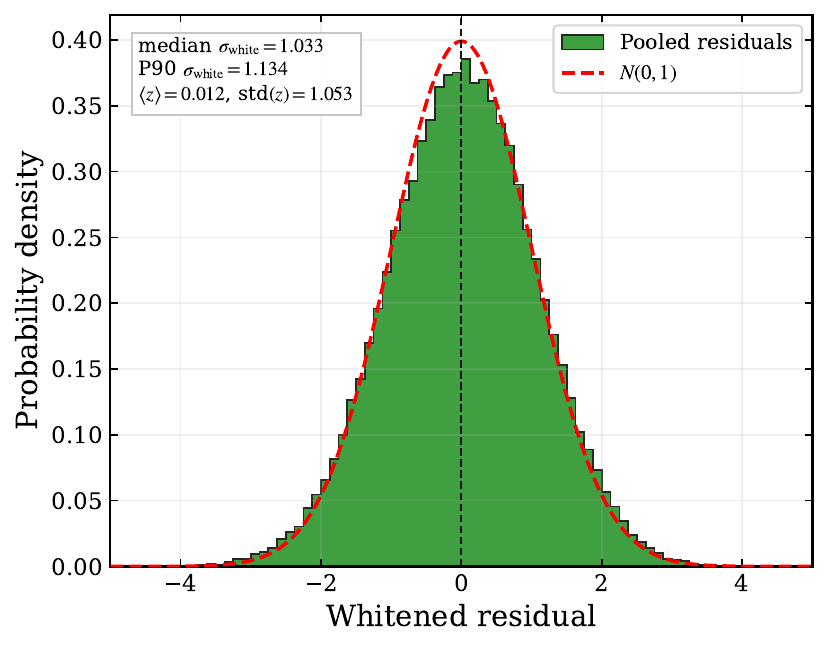}
\end{minipage}

\caption{
Validation of the fiducial background model.
(a) Distribution of the whitened-residual standard deviation $\sigma_{\rm white}$ in different energy bands and instruments.
Green points denote individual $(X,r)$ spectra, box plots show the distribution across rings, and the red dashed line indicates the ideal value $\sigma_{\rm white}=1$.
(b) Distribution of the whitened residuals pooled over all energy bands, rings, and instruments.
The green histogram shows the probability density of the pooled residuals, while the red dashed curve shows the standard normal distribution $N(0,1)$.
The pooled residuals give median $\sigma_{\rm white}=1.033$, $P_{90}(\sigma_{\rm white})=1.134$, $\langle z\rangle=0.012$, and ${\rm std}(z)=1.053$.
}
\label{fig:bg_validation}
\end{figure*}

Fig.~\ref{fig:bg_validation} shows the stability tests of the three-band background model for all MOS and PN rings. Fig.~\ref{fig:bg_validation}(a) presents the distribution of $\sigma_{\rm white}$ for all data sets. The results are clustered around $\sigma_{\rm white}=1$, indicating that the scale of the model residuals is broadly consistent with the estimated covariance. A small number of rings show relatively large local residuals, but they do not dominate the overall distribution of the fit-quality statistic. In particular, Fig.~\ref{fig:bg_validation}(b) shows that the whitened residual distribution pooled over all energy bands, rings, and instruments is close to the standard normal distribution $N(0,1)$. Quantitatively, the pooled test gives
\begin{equation}
{\rm median}\,\sigma_{\rm white}=1.033,
\qquad
{\rm P90}\,\sigma_{\rm white}=1.134,
\end{equation}
and
\begin{equation}
\langle z\rangle = 0.012,
\qquad
{\rm std}(z)=1.053.
\end{equation}
These results indicate that the combined fiducial background model provides a stable statistical description of the current data set.

\section{result and discuss}
\label{section_result}

\subsection{Likelihood construction and validation}
\label{subsection_likelihood}

After constructing and validating the fiducial background model, we perform an independent narrow-line search for each ALP mass point. Since the XMM-BSO data are count-rate spectra obtained by stacking a large number of exposures, and the number of counts in each energy bin is large, we use a Gaussian likelihood to describe the spectral fit.

For a given energy band $b$, instrument $X=\mathrm{MOS},\mathrm{PN}$, ring $r$, and energy bin $k$, we denote the observed count-rate spectrum by $d_{X,r,k}$ and its statistical uncertainty by $\sigma_{X,r,k}$. For a given ALP mass $m_a$, the model expectation is written as
\begin{equation}
\mu_{X,r,k}(g_{a\gamma\gamma},m_a,\boldsymbol\theta)
=
B_{X,r,k}(\boldsymbol\theta)
+
S_{X,r,k}(g_{a\gamma\gamma},m_a),
\end{equation}
where \(B_{X,r,k}(\boldsymbol\theta)\) is the background model family constructed in the previous section, and \(\boldsymbol\theta\) denotes the corresponding nuisance parameters, including the continuum mean-function coefficients, line amplitudes, and local nuisance modes. The GP covariance is fixed to the result obtained from the background-only fit described in Sec.~\ref{sec:bg_validation}. The signal template \(S_{X,r,k}(g_{a\gamma\gamma},m_a)\) is computed directly for each trial value of \(g_{a\gamma\gamma}\) and \(m_a\), following Eq.~\eqref{eq:xmmbso_signal_template}. This direct parameterization automatically includes the finite-lifetime and bound-state accumulation effects in the signal normalization.

For an ALP with mass $m_a$, the profile likelihood for the signal search is then written as
\begin{widetext}
\begin{equation}
-2\ln \mathcal L(g_{a\gamma\gamma},\boldsymbol\theta;m_a)
=
\sum_{X,r}
\left[
\mathbf d_{X,r}
-
\boldsymbol\mu_{X,r}(g_{a\gamma\gamma},m_a,\boldsymbol\theta)
\right]^T
\left(\mathbf C^{(b)}_{X,r}\right)^{-1}
\left[
\mathbf d_{X,r}
-
\boldsymbol\mu_{X,r}(g_{a\gamma\gamma},m_a,\boldsymbol\theta)
\right]
+
\sum_{X,r}
\ln\det \mathbf C^{(b)}_{X,r}
+
{\rm const}.
\end{equation}
\end{widetext}
Here \(\mathbf C^{(b)}_{X,r}\) is the data covariance matrix defined in Eq.~\eqref{eq:cov_total}. Since \(\mathbf C^{(b)}_{X,r}\) is fixed during the ALP mass scan, the \(\ln\det \mathbf C^{(b)}_{X,r}\) term does not affect the best-fit coupling \(g_{a\gamma\gamma}\) or the profiled nuisance parameters. We nevertheless keep it explicitly in the equation to display the full form of the Gaussian likelihood.

We further define the profile-likelihood ratio \cite{Wilks:1938dza,Cowan:2010js}
\begin{equation}
q(g_{a\gamma\gamma})
=
-2\ln
\frac{
\mathcal L(g_{a\gamma\gamma},\widehat{\boldsymbol\theta}_{g};m_a)
}{
\mathcal L(\widehat g_{a\gamma\gamma},\widehat{\boldsymbol\theta};m_a)
},
\end{equation}
where \(\widehat{\boldsymbol\theta}_{g}\) denotes the nuisance-parameter best fit at fixed \(g_{a\gamma\gamma}\), and \((\widehat g_{a\gamma\gamma},\widehat{\boldsymbol\theta})\) denotes the global best fit. When deriving upper limits, we impose the physical condition \(g_{a\gamma\gamma}\geq0\) and use the one-sided \(95\%\) criterion
\begin{equation}
q(g_{a\gamma\gamma}^{95})=2.71
\end{equation}
to define the upper limit \(g_{a\gamma\gamma}^{95}(m_a)\).

\subsection{Limit on axion-photon coupling}

We first consider the case in which ALPs couple only to photons. In this scenario, ALPs are produced in the solar interior through the Primakoff process and photon coalescence, become gravitationally bound and accumulate around the Sun as a solar-basin population, and subsequently decay into photons that can be detected by X-ray telescopes \cite{VanTilburg:2020jvl,DeRocco:2022jyq,Beaufort:2023zuj}. The stacked blank-sky spectra constructed from all-sky XMM-Newton observations therefore provide a probe of the ALP--photon coupling.

\begin{figure}
    \centering
    \includegraphics[width=1.0\linewidth]{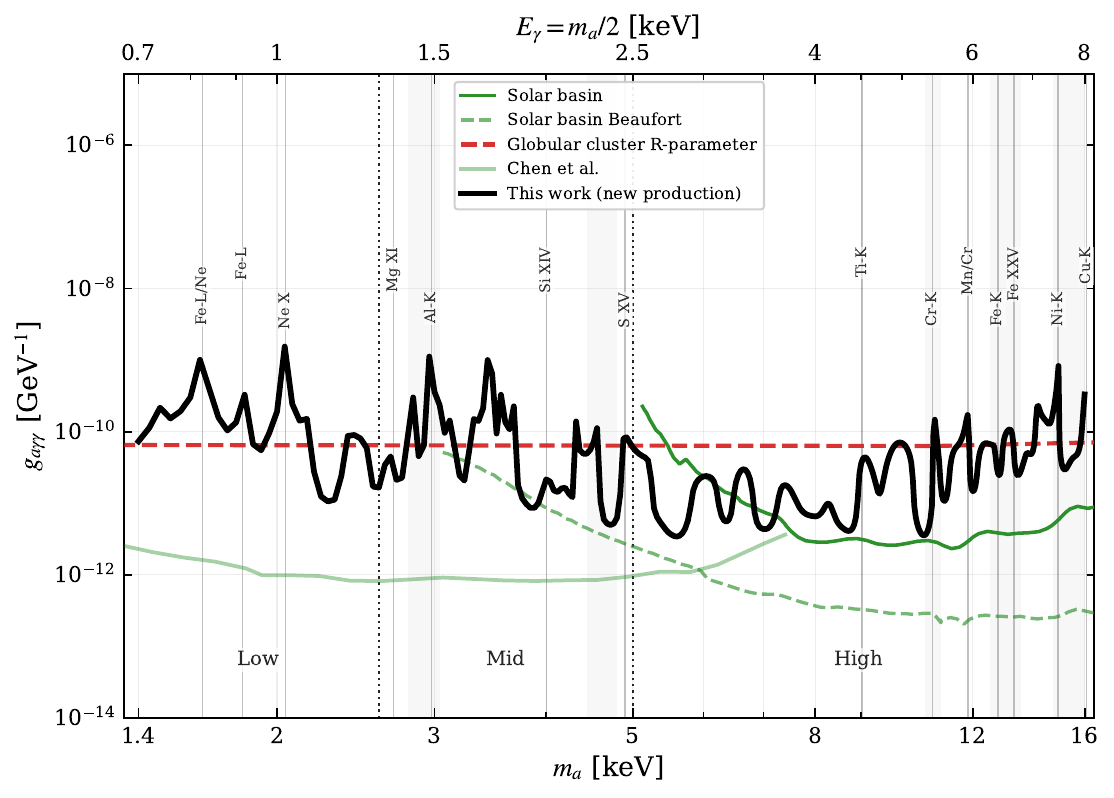}
    \caption{ The $95\%$ confidence-level upper limits on the ALP--photon coupling $g_{a\gamma\gamma}$. The lower horizontal axis shows the ALP mass $m_a$, while the upper horizontal axis shows the corresponding decay-photon energy $E_\gamma=m_a/2$. The black solid curve gives the limit obtained in this work. The green curves show existing solar-basin or X-ray-related constraints \cite{DeRocco:2022jyq,Beaufort:2023zuj,Chen:2024ekh}, and the red dashed curve shows the globular-cluster $R$-parameter constraint\cite{Ayala:2014pea,AxionLimits}. The vertical black dashed lines mark the boundaries between the low-, mid-, and high-energy background models. The gray vertical lines indicate the positions of several known astrophysical lines or instrumental fluorescence features that have a sizable impact on the limit. } \label{fig:gagg_limit}
\end{figure}

We find no significant evidence for an unidentified narrow-line signal, and therefore set upper limits on the ALP--photon coupling. Fig.~\ref{fig:gagg_limit} shows the resulting $95\%$ confidence-level upper limits. The lower horizontal axis denotes the ALP mass $m_a$, while the upper horizontal axis denotes the corresponding decay-photon energy $E_\gamma=m_a/2$. The line search covers the photon-energy range $E_\gamma=0.7\text{--}8~{\rm keV}$, corresponding to the ALP mass range $m_a=1.4\text{--}16~{\rm keV}$. The black solid curve shows the observed limit obtained in this work from the XMM-BSO data. Over most of the mass range $m_a=1.4\text{--}16~{\rm keV}$, our limit lies at the level of $g_{a\gamma\gamma}\sim10^{-11}\text{--}10^{-10}~{\rm GeV}^{-1}$. The red dashed curve shows the globular-cluster constraint from the $R$-parameter \cite{Ayala:2014pea}, while the green curves show existing constraints from solar-basin or stellar X-ray observations \cite{DeRocco:2022jyq,Beaufort:2023zuj,Chen:2024ekh}. Our result is comparable to the globular-cluster bound and becomes stronger in several high-energy windows. Compared with existing stellar X-ray constraints, our analysis uses the XMM-BSO stacked blank-sky spectra and an independently constructed line-search background model, and therefore provides a complementary constraint on keV-mass ALPs.

The gray vertical lines and shaded regions in Fig.~\ref{fig:gagg_limit} mark the positions of several known astrophysical lines and instrumental fluorescence features. The weakening of the limit curve in several local regions coincides with these line positions, indicating that the loss of sensitivity is mainly driven by degeneracies with narrow-line background components. This behavior is expected in narrow-line searches when the signal template overlaps with known background-line templates \cite{Foster:2021ngm,Jeltema:2015mee,Dessert:2018qih}. We emphasize, however, that not every background line produces the same level of degradation. The size of the degeneracy depends on the actual line intensity, the instrumental response, the stacking procedure, the local continuum uncertainty, and the presence of nearby nuisance-basis components.

\subsection{Impact of the Ring-Level XMM-BSO Response Approximation}

As discussed above, the public XMM-BSO data products provide only spectra and instrumental responses stacked at the instrument/ring level. We therefore adopted a ring-averaged response approximation when constructing the ALP signal templates. The validity of this approximation depends on the spatial, observing-mode, and energy dependence of the instrumental response. To assess its impact on the final limits, we perform response-stress tests on the ring-level detector response. We consider two classes of perturbations: response-shape variations and response-normalization variations.

The first class is a response-shape test, designed to evaluate the impact of changes in the response-convolved line shape. For MOS and PN separately, we construct a median response shape from the normalized response shapes of the 30 rings. We then select the ring response with the smallest cosine similarity to the median shape as a conservative shape-stress case.

The second class is a response-normalization test, designed to assess the impact of an overall change in the effective-area scale. The official EPIC calibration status note quotes an absolute effective-area uncertainty of approximately $\pm10\%$ \cite{Read:2014vga}. We therefore adopt a $\pm10\%$ normalization variation as the response-normalization stress test.

We find that changing the ring-level response shape leads to a maximum relative change of only $3.5\times10^{-4}$ in the coupling limit. This indicates that variations in the response-convolved line shape have a negligible impact on the final limit curve. Under the $\pm10\%$ absolute effective-area calibration envelope, the maximum change in $g_{95}$ is approximately $5.4\%$. We therefore regard the ring-level response approximation used in this work as a controlled systematic uncertainty. A fully self-consistent treatment of the instrumental response would require reprocessing all XMM-Newton observations entering the stacked data set. Such a complete data-reduction and response-construction pipeline is beyond the scope of the present work, and is left for future investigation.

\section{Conclusions}
\label{conclusion}

In this work, we have presented a search strategy for narrow decay-line signals from the solar gravitationally bound ALP basin using the XMM-BSO data. We derived the count templates for the decay signal in the XMM-BSO stacked spectra and constructed a piecewise spectral background model over the $0.7\text{--}8.0~{\rm keV}$ photon-energy range. We then built a profile likelihood to search for the ALP signal. In the mass range $m_a=1.4\text{--}16~{\rm keV}$, we obtained $95\%$ confidence-level limits on $g_{a\gamma\gamma}$, with typical sensitivities of $g_{a\gamma\gamma}\sim10^{-11}\text{--}10^{-10}~{\rm GeV}^{-1}$. These limits are comparable to the globular-cluster bound, although they are generally weaker than the strongest existing constraints from stellar X-ray observations. Nevertheless, our result provides a cross-check of keV-mass ALP decay from the solar bound population.

Our analysis further shows that the local weakening of the limit curve mainly occurs near known regions of background complexity. This is caused by degeneracies between the trial ALP decay line and strong astrophysical lines or instrumental fluorescence features: when the signal line lies close to such features, the background nuisance templates can absorb part of the line-like information and reduce the distinguishability of an additional ALP signal. Through response-shape and response-normalization stress tests, we found that the ring-level response approximation adopted in this work can be treated as a controlled systematic uncertainty.

Current and future X-ray missions, such as \textit{XRISM} \cite{2020arXiv200304962X}, \textit{Athena} \cite{2013arXiv1306.2307N}, and \textit{eXTP} \cite{2025SCPMA..6819502Z}, will further improve searches for narrow X-ray lines through improved spectral resolution, larger effective exposure, and more flexible observing strategies. For solar-basin ALP decay, the advantages of future instruments will come not only from improved energy resolution, but also from better separation of background lines, more stable instrumental-response modeling, and more flexible choices of observing geometry. This work demonstrates the feasibility of using blank-sky stacked spectra to search for decay lines from solar-bound ALPs, and provides a scalable analysis framework for future searches with higher-quality X-ray data.

\acknowledgments
This work is supported by the National Key R\&D Program of China (Grants No. 2022YFF0503304), the National Natural Science Foundation of China (12373002, 12220101003, 11773075) and the Youth Innovation Promotion Association of Chinese Academy of Sciences (Grant No. 2016288).


\bibliographystyle{apsrev4-2}
\bibliography{refs}

\newpage

\appendix
\setcounter{equation}{0}
\setcounter{figure}{0}

\section{Details of XMM-BSO Background Model}
\label{sec:appendix_background_details}

\subsection{Low-Band Model}

The low-energy band, $0.7\text{--}1.3~{\rm keV}$, covers a relatively complex part of the soft X-ray spectrum. This energy range contains contributions from astrophysical line complexes, including the Fe-L complex and Ne-like line features \cite{1995ApJ...438L.115L,2001ApJ...556L..91S}, and may also contain local residual structures induced by the stacking procedure, instrumental response, and background treatment of XMM-Newton EPIC data \cite{2007A&A...464.1155C,2008A&A...478..575K,Foster:2021ngm}. In this subsection, we do not attempt to assign a physical origin to every low-energy residual feature. Instead, we use an empirical line library obtained from background-only residual scans to describe stable line-like structures in the data.

Specifically, the empirical line-template library used in the low-energy band is
\begin{equation}
\mathcal L_{\rm low}
=
\{0.727,\ 0.826,\ 0.915,\ 1.022,\ 1.095\}~{\rm keV}.
\end{equation}
The templates in the $0.727\text{--}0.915~{\rm keV}$ range effectively cover Fe-L/Ne-like soft X-ray residual structures, while the $1.022~{\rm keV}$ template lies close to the Ne X Ly$\alpha$ line region. The $1.095~{\rm keV}$ template is included to absorb a stable local residual that remains after the background-only pre-fits. We emphasize that these labels are empirical: the templates are introduced to describe stable background structures, and not every template is assumed to correspond to a uniquely identified astrophysical or instrumental transition.

Using only fixed-center line templates still leaves residuals resembling small centroid shifts or asymmetric composite-line shapes in some rings. We therefore introduce shift templates for the following three relatively strong low-energy features:
\begin{equation}
\mathcal L_{\rm shift}
=
\{0.826,\ 0.915,\ 1.022\}~{\rm keV}.
\end{equation}
For a line centered at $E_\ell$, the shift template is defined as
\begin{equation}
S_{X,r,\ell,k}
=
\frac{
L_{X,r,k}(E_\ell+\Delta E)
-
L_{X,r,k}(E_\ell-\Delta E)
}{2},
\end{equation}
where $\Delta E=0.005~{\rm keV}$.
These shift templates do not represent additional independent spectral lines. Instead, they describe small centroid shifts of the fixed line templates or asymmetric composite-line shapes under the finite energy resolution of the instrument.

The fiducial low-band background model is therefore
\begin{widetext}
\begin{equation}
B^{\rm low}_{X,r,k}
=
p_{X,r,k}
+
g_{X,r,k}
+
\sum_{\ell\in\mathcal L_{\rm low}}
a_{X,r,\ell}
L_{X,r,\ell,k}
+
\sum_{\ell\in\mathcal L_{\rm shift}}
b_{X,r,\ell}
S_{X,r,\ell,k}.
\end{equation}
\end{widetext}
Here $b_{X,r,\ell}$ are local shape nuisance parameters associated with the shift templates. The continuum coefficients $c_{X,r,n}$, the ordinary line amplitudes $a_{X,r,\ell}$, and the shift-template amplitudes $b_{X,r,\ell}$ are all treated as local nuisance parameters and are profiled independently in each $(X,r)$ data set.

\subsection{Mid-Band Model}

The mid-energy band, $1.3\text{--}2.5~{\rm keV}$, contains several soft X-ray atomic line features as well as strong instrumental fluorescence lines. In contrast to the low-energy band, the residual structures in this band can be more directly associated with Mg/Si/S line features in thermal astrophysical plasmas \cite{2001ApJ...556L..91S,2012ApJ...756..128F} and with the Al-K fluorescence feature. We therefore do not use a purely empirical line library in this band. Instead, we decompose the mid-band line background into two parts: a set of identified astrophysical line templates and a local instrumental-shape correction around the Al-K complex.

Because astrophysical line emission can have a non-trivial distribution as a function of Galactic-Center angle, we use the profile-linked physical-line parameterization described above. The fiducial linked physical-line candidates are
\begin{equation}
\mathcal L_{\rm phys3}
=
\left\{
{\rm Mg\,XI}\ 1.350,\ 
{\rm Si\,XIV}\ 2.000,\ 
{\rm S\,XV}\ 2.450
\right\}\ {\rm keV}.
\end{equation}
These templates represent the stable Mg-, Si-, and S-line structures observed in the mid-energy residuals. Their ring profiles are determined from background-only pre-fits and then kept fixed in the ALP search, leaving only one linked amplitude parameter for each line.

In addition to these astrophysical line templates, the mid-energy band contains an important local instrumental feature: the Al-K complex centered around $E_{\rm AlK}\simeq1.49~{\rm keV}$. This energy is close to the Mg XII Ly$\alpha$ line region at $1.470~{\rm keV}$, and the Al-K instrumental fluorescence feature can leave residual structures in ring-stacked spectra if its line wing, composite-line asymmetry, or local curvature is not perfectly captured by a single fixed line template \cite{2007A&A...464.1155C,2008A&A...478..575K,Foster:2021ngm}. We therefore introduce a dedicated local shape-correction template for the Al-K complex. Denoting this fixed curvature template by $C^{\rm AlK}_{X,r,k}$, the local Al-K component is written as
\begin{equation}
B^{\rm AlK}_{X,r,k}
=
a^{\rm AlK}_{X,r}
L^{\rm AlK}_{X,r,k}
+
b^{\rm AlK}_{X,r}
C^{\rm AlK}_{X,r,k}.
\end{equation}
Here $a^{\rm AlK}_{X,r}$ and $b^{\rm AlK}_{X,r}$ are local nuisance parameters profiled independently in each $(X,r)$ data set. The template $L^{\rm AlK}_{X,r,k}\equiv L_{X,r,k}(E_{\rm AlK})$ denotes the main Al-K line template. The template $C^{\rm AlK}_{X,r,k}$ is a phenomenological local curvature mode around the Al-K complex, proportional to $\partial_E^2 L_{X,r,k}(E)|_{E=E_{\rm AlK}}$. Numerically, we construct it from a second-order finite difference of the same Al-K response template,
\begin{widetext}
\begin{equation}
C^{\rm AlK}_{X,r,k}
\equiv
L_{X,r,k}(E_{\rm AlK}+\Delta_{\rm AlK})
+
L_{X,r,k}(E_{\rm AlK}-\Delta_{\rm AlK})
-
2L_{X,r,k}(E_{\rm AlK}),
\end{equation}
with $\Delta_{\rm AlK}=0.010~{\rm keV}$. Since the overall normalization of this curvature mode is absorbed by the free coefficient $b^{\rm AlK}_{X,r}$, we do not divide the finite difference by $\Delta_{\rm AlK}^2$ in the numerical implementation.

Combining these components, the fiducial mid-band background model is

\begin{equation}
B^{\rm mid}_{X,r,k}
=
p_{X,r,k}
+
g_{X,r,k}
+
\sum_{\ell\in\mathcal L_{\rm phys3}}
A_{X,\ell}
P_{X,r,\ell}
L_{X,r,\ell,k}
+
a^{\rm AlK}_{X,r}
L^{\rm AlK}_{X,r,k}
+
b^{\rm AlK}_{X,r}
C^{\rm AlK}_{X,r,k}.
\end{equation}
\end{widetext}
Here $p_{X,r,k}$ is the low-order continuum mean function, $g_{X,r,k}$ is the GP residual, the profile-linked physical templates describe the Mg XI, Si XIV, and S XV line structures, and the Al-K local templates describe the instrumental line-shape mismatch around $1.49~{\rm keV}$. This construction keeps the model flexible enough to account for the dominant stable residual structures in the mid-energy band, while limiting the number of local nuisance modes that could absorb a narrow ALP decay-line signal.

\subsection{High-Band Model}

The high-energy band, $2.5\text{--}8.0~{\rm keV}$, contains the largest number of instrumental fluorescence features and high-energy astrophysical lines in the XMM-BSO spectra. Compared with the low- and mid-energy bands, this band includes more background lines, with several features located close to each other in energy. We therefore adopt the high-band line atlas used by Foster et al.~\cite{Foster:2021ngm}, and treat these lines uniformly with the profile-linked parameterization described above. The astrophysical line identifications are motivated by standard thermal-plasma line complexes \cite{2001ApJ...556L..91S,2012ApJ...756..128F}, while the instrumental fluorescence features are guided by the known XMM-Newton EPIC background components \cite{2007A&A...464.1155C,2008A&A...478..575K}.

The high-band line library used in this work is

\begin{equation}
\mathcal L_{\rm high}
=
\left\{
\begin{array}{l}
{\rm S\,XVI}\ 2.62,\ 
{\rm Ar\,XVII}\ 3.12,\ \\
{\rm Ar\,XVIII}\ 3.32,\ 
{\rm Ca\,XIX}\ 3.90,\ \\
{\rm Ca\,XX}\ 4.11,
{\rm Ti\,K}\ 4.51,\ \\
{\rm Ti\,K\beta}\ 4.93,\ 
{\rm Cr\,K}\ 5.41,\ \\
{\rm Mn\,K}\ 5.90,\ 
{\rm Cr\,K\beta}\ 5.95,\\
{\rm Fe\,K}\ 6.40,\ 
{\rm Fe\,XXV}\ 6.67,\ \\
{\rm Fe\,XXVI}\ 6.97,\ 
{\rm Ni\,K}\ 7.47,\ \\
{\rm Cu\,K}\ 8.04
\end{array}
\right\}\ {\rm keV}.
\end{equation}

The labels, such as ${\rm S\,XVI}$, are used to indicate the likely origin of the corresponding spectral features. The S/Ar/Ca-like lines are more naturally associated with astrophysical thermal-plasma emission, whereas the Ti/Cr/Mn-like features are more closely related to instrumental or particle-background-induced fluorescence structures. The Fe K feature around $6.40~{\rm keV}$ can include neutral or low-ionization Fe K$\alpha$ emission from the sky background, and may also receive contributions from Fe-like instrumental or particle-background fluorescence. The Ni/Cu-like structures around $7.47\text{--}8.04~{\rm keV}$ lie close to the upper boundary of the high-energy band, where their impact can arise from fixed narrow lines, instrumental-response tails, and the truncation of the fitting window.

The fiducial high-band background model is therefore written as
\begin{equation}
B^{\rm high}_{X,r,k}
=
p_{X,r,k}
+
g_{X,r,k}
+
\sum_{\ell\in\mathcal L_{\rm high}}
A^{\rm high}_{X,\ell}
P^{\rm high}_{X,r,\ell}
L^{\rm high}_{X,r,\ell,k}.
\end{equation}
Here $p_{X,r,k}$ and $g_{X,r,k}$ describe the smooth continuum and GP residual, respectively, while the linked high-band line templates describe the stable narrow-line structures identified in the XMM-BSO high-energy spectra. The ring profiles $P^{\rm high}_{X,r,\ell}$ are obtained from background-only pre-fits and kept fixed in the ALP search, with only the linked amplitudes $A^{\rm high}_{X,\ell}$ profiled.

\end{document}